# Influence of fluctuations on the superfluid density in low-temperature technologies

Iogann Tolbatov

*Physics and Engineering Department, Kuban State University, Krasnodar, Russia*

*(talbot1038@mail.ru)*

*In this article the Ginzburg-Landau theory ideas are considered in their application to the description of fluctuations influence on the superfluid density in superconductor. The conclusion about the availability of two incompatible mathematical definitions of the superfluid density is made. That is why, it is suggested not to consider the fluctuating part of order parameter, while calculating the superconductor thermodynamical characteristics in the Mean Field Approximation.*

The fluctuation Cooper pairs appearence in the vicinity of the matter transition to the superconducting state produces an effect, which changes the magnetic field applied to the superconductor expulsion character. This problem was studied in the works [Buzdin and Vuychich, Ref. 1], [Glazman and Koshlev, Ref. 2]. In the superconductor simple electrodynamics frameworks the expressions

$$\bar{j} = -n_s \frac{2e^2}{m} \bar{A} = -\frac{1}{4\pi\lambda^2(T)} \bar{A}, \quad \lambda^2(T) = \frac{m}{8\pi n_s e^2} \tag{1}$$

relate the superconducting current density $j$, the penetration depth $\lambda(T)$ with the superfluid density $n_s$ ("the superconducting electrons density" in the London theory terms [F. London and H. London, Ref. 3]).

The Ginzburg-Landau theory permits to calculate the temperature dependence of magnetic field penetration depth in the superconductor volume $\lambda(T)$ in the critical temperature vicinity and another superconducting sample characteristics. At the same time the supefluid density is identified with the order parameter average value

$$n_s = \tilde{\Psi}^2 = -\frac{a}{b}, \tag{2}$$

and at the temperatures below the critical $T_C$ the supercurrent is written in the form [Buzdin and Vuychich, Ref. 1], [Glazman and Koshlev, Ref. 2]:

$$\langle \bar{j}_{fl} \rangle = -\frac{2e^2}{m}\bar{A}\left[\tilde{\Psi}^2 - 2\sum_{\bar{k}}\frac{T}{\left(2\alpha T_C |\varepsilon| + \frac{\bar{k}^2}{4m}\right)}\right], \tag{3}$$

where $\varepsilon$ is the reduced temperature.

The corresponding to the expression (3) superfluid density definition, while one takes into account the fluctuations and (1), turns out to be:

$$n_s(T) = \frac{\alpha}{b}\left[T_C|\varepsilon| - \frac{2b}{\alpha^2}\sum_{\bar{k}}\frac{1}{\left(2|\varepsilon| + \frac{\bar{k}^2}{4m}\alpha T_C\right)}\right]. \tag{4}$$

In the Mean Field Approximation the superfluid density equals the average square of the coordinate depending order parameter [Varlamov and Larkin, Ref. 4]:

$$\langle |\Psi(\bar{r})|^2 \rangle = \tilde{\Psi}^2 = n_s(T). \tag{5}$$

Separating equilibrium and fluctuating parts of the order parameter $\Psi(r)$ and taking into account that $\langle |\Psi(\bar{r})|^2 \rangle = \tilde{\Psi}^2 + \langle \psi_r^2 \rangle + \langle \psi_i^2 \rangle + 2\tilde{\Psi}\langle \psi_r \rangle$, we find the following expression:

$$\langle |\Psi(\bar{r})|^2 \rangle = \tilde{\Psi}^2 - 2\langle \psi_r^2 \rangle. \tag{6}$$

Equilibrium and fluctuating parts of the order parameter accordingly are equal to:

$$\langle \psi_r^2 \rangle = \frac{T}{2}\sum_{\bar{k}}\frac{1}{2\alpha T_C|\varepsilon| + \frac{\bar{k}^2}{4m}} \quad \text{and} \quad \langle \psi_i^2 \rangle = \frac{T}{2}\sum_{\bar{k}}\frac{1}{\frac{\bar{k}^2}{4m}}.$$

It follows from the dependence (6) that:

$$\left\langle |\Psi(\bar{r})|^2 \right\rangle = \frac{\alpha}{b}\left[T_C|\varepsilon| - \frac{b}{\alpha^2}\sum_{\bar{k}}\frac{1}{\left(2|\varepsilon| + \frac{\bar{k}^2}{4m}\alpha T_C\right)}\right]. \tag{7}$$

So we conclude – the fluctuation consideration leads to the difference between $\left\langle |\Psi(\bar{r})|^2 \right\rangle$ and $n_s(T)$ already in the first order.

In cases of the effective dimensionalities $2D$ and $3D$ the summation over momenta in (4) formally diverges at the upper limit [Varlamov and Larkin, Ref. 4]. This ultra-violet catastrophe is related with the Ginzburg-Landau functional applicability limits violation for $|\bar{k}| > \xi^{-1}$, and it is cut off at $\xi \cdot |\bar{k}| = C_{(D)} \sim 1$.

Our purpose is to show that the divergence of the sum over momenta in the expression (4) occurs, because of the fluctuation consideration in the Mean Field Approximation. The above-mentioned ultra-violet catastrophe study enabled us to conclude that it is not the consequence of the sum divergence at the upper limit, but it appears at the earlier stage, since the superfluid density in the Ginzburg-Landau formalism has the two incompatible mathematical definitions.

For to accomplish the above-stated purpose we write down (4) and (7) in the form:

$$n_s(T) = \frac{\alpha T_C|\varepsilon|}{b} - \frac{2}{\alpha}\sum_{\bar{k}}\frac{1}{\left(2|\varepsilon| + \frac{\bar{k}^2}{4m}\alpha T_C\right)}, \tag{8}$$

$$\left\langle |\Psi(\bar{r})|^2 \right\rangle = \frac{\alpha T_C|\varepsilon|}{b} - \frac{1}{\alpha}\sum_{\bar{k}}\frac{1}{\left(2|\varepsilon| + \frac{\bar{k}^2}{4m}\alpha T_C\right)}. \tag{9}$$

From the expressions (8) and (9) we find:

$$n_s(T) = \left\langle |\Psi(\bar{r})|^2 \right\rangle - \frac{1}{\alpha}\sum_{\bar{k}}\frac{1}{\left(2|\varepsilon| + \frac{\bar{k}^2}{4m}\alpha T_C\right)}. \tag{10}$$

Investigating the dependencies (5), (6), and (10), we have to notice that:

$$\left\langle \psi_r^2 \right\rangle = \frac{1}{2\alpha}\sum_{\bar{k}}\frac{1}{\left(2|\varepsilon| + \frac{\bar{k}^2}{4m}\alpha T_C\right)}. \tag{11}$$

From expressions (10) and (11) we conclude that:

$$n_s(T) = \langle |\Psi(\vec{r})|^2 \rangle - 2\langle \psi_r^2 \rangle. \tag{12}$$

After the substitution of (5) in (6) we find:

$$\langle |\Psi(\vec{r})|^2 \rangle + 2\langle \psi_r^2 \rangle = \tilde{\Psi}^2 = n_s(T). \tag{13}$$

We compare (13) with the expression (12):

$$\langle |\Psi(\vec{r})|^2 \rangle + 2\langle \psi_r^2 \rangle = \langle |\Psi(\vec{r})|^2 \rangle - 2\langle \psi_r^2 \rangle. \tag{14}$$

The found equation is valid only under condition $\langle \psi_r^2 \rangle = 0$, $\dfrac{1}{2\alpha}\sum_{\vec{k}} \dfrac{1}{\left(2|\varepsilon| + \dfrac{\vec{k}^2}{4m}\alpha T_C\right)} = 0$. And this statement is true with the consideration of any order fluctuations.

As the result of the carried out investigation we obtain the following inferences:

(1) The fluctuating part average square is equal to zero. That is why, while calculating superconductor thermodynamical characteristics in the Mean Field Approximation, one has not to take into account the fluctuating part of order parameter.

(2) One has not to use the expression $n_s(T) = \dfrac{\alpha}{b}\left[T_C|\varepsilon| - \dfrac{2b}{\alpha^2}\sum_{\vec{k}} \dfrac{1}{\left(2|\varepsilon| + \dfrac{\vec{k}^2}{4m}\alpha T_C\right)}\right]$

in the description of the fluctuations influence on the superfluid density and critical temperature, not because of the sum over momenta divergence at the upper limit.

The cause of the (1) and (2) inferences is in the availability of two incompatible mathematical definitions of the superfluid density, which both are used in the Ginzburg-Landau methodology.

References
[1] A. I. Buzdin and B. Vuychich, Modern Phys. Lett. B **4**, 485 (1990).